\documentclass[aps,prd,preprint,superscriptaddress,tightenlines,nofootinbib,showpacs]{revtex4}

\usepackage{graphicx}
\usepackage{dcolumn}
\usepackage{bm}

\def\mtiny{\vrule width 0pt}
\def\mrm#1{\mathrm{#1}}
\def\DZ{\relax\ifmmode{D^0}\else{$\mrm{D}^{\mrm{0}}$}\fi}
\def\DONE{\relax\ifmmode{D_1}\else{$\mrm{D}_{\mrm{1}}$}\fi}
\def\DTWO{\relax\ifmmode{D_2}\else{$\mrm{D}_{\mrm{2}}$}\fi}
\def\KZ{\relax\ifmmode{K^0}\else{$\mrm{K}^{\mrm{0}}$}\fi}
\def\KSHO{\relax\ifmmode{K_{\rm S}}\else{$\mrm{K}_{\mrm{S}}$}\fi}
\def\KLON{\relax\ifmmode{K_{\rm L}}\else{$\mrm{K}_{\mrm{L}}$}\fi}
\def\BZ{\relax\ifmmode{B^0}\else{$\mrm{B}^{\mrm{0}}$}\fi}\def\BZp{\relax\ifmmode{B^0}\else{$\mrm{B}^{\mrm{0}}$}\fi}
\def\BZS{\relax\ifmmode{B^0_s}\else{$\mrm{B}^{\mrm{0}_s}$}\fi}
\def\DZS{\relax\ifmmode{D^{*+}}\else{$\mrm{D}^{\mrm{*+}}$}\fi}
\def\DZB{\relax\ifmmode{\overline{D}\mtiny^0}
        \else{$\overline{\mrm{D}}\mtiny^{\mrm{0}}$}\fi}
\def\KZB{\relax\ifmmode{\overline{K}\mtiny^0}
        \else{$\overline{\mrm{K}}\mtiny^{\mrm{0}}$}\fi}
\def\BZB{\relax\ifmmode{\overline{B}\mtiny^0}
        \else{$\overline{\mrm{B}}\mtiny^{\mrm{0}}$}\fi}
\def\BZBp{\relax\ifmmode{\overline{B}\mtiny^0}
        \else{$\overline{\mrm{B}}\mtiny^{\mrm{0}}$}\fi}
\def\BZBS{\relax\ifmmode{\overline{B}\mtiny^0_s}
        \else{$\overline{\mrm{B}}\mtiny^{\mrm{0}_s}$}\fi}
\def\DZC{\relax\ifmmode{\overline{D}\mtiny^0}
        \else{$\overline{\mrm{D}}\mtiny^{\mrm{0}}$}\fi}
\begin{document}

\preprint{CLNS 05/1923}       

\title{\boldmath Time-Independent Measurements of $D^0$-$\DZB$ Mixing and
Relative Strong Phases Using Quantum Correlations}

\author{D.~M.~Asner}
\affiliation{Carleton University, Ottawa, Ontario, Canada K1S 5B6}
\author{W.~M.~Sun}
\affiliation{Cornell University, Ithaca, New York 14853}

\date{November 6, 2007}

\begin{abstract}
Due to quantum correlations in the $C$-odd and $C$-even $D^0\DZB$ pairs
produced in the reactions $e^+e^-\to D^0\DZB(n\pi^0)$ and
$e^+e^-\to D^0\DZB\gamma(n\pi^0)$, respectively, the time-integrated
$D^0\DZB$ decay rates are sensitive to interference between amplitudes for
indistinguishable final states.  The size of this interference is governed by
the relevant amplitude ratios and can include contributions from
$D^0$-$\DZB$ mixing.  We present a method for simultaneously measuring
the magnitudes and phases of these amplitude ratios and
searching for $D^0$-$\DZB$ mixing.  We make use of fully- and
partially-reconstructed $D^0\DZB$ pairs in both $C$ eigenstates, and we
estimate experimental sensitivities based on a plausible charm factory dataset.
Similar analyses can be applied to coherent $K^0\KZB$, $B^0\BZB$, or
$B_s^0\BZBS$ pairs.
\end{abstract}

\maketitle

\section{Introduction}\label{sec:intro}

Studies of the evolution of a $\KZ$ or $\BZ$ into the respective
anti-particle, a $\KZB$ or $\BZB$~\cite{gmp},
have guided the form and content of the Standard Model and permitted
useful estimates of the masses of the charm~\cite{goodetal,glr}
and top quark~\cite{Albrecht,rosner87} prior to their direct observation.
Neutral flavor oscillation in the $D$ meson system is highly suppressed
within the Standard Model and, thus, with current experimental sensitivity,
searches for $\DZ$-$\DZB$ mixing constitute a search for new physics.
In addition, improving constraints on charm mixing is important for
elucidating the origin of $CP$ violation in the bottom sector.

The time evolution of the $\DZ$-$\DZB$ system is 
described by the Schr\"odinger equation
\begin{equation}
i \frac{\partial}{\partial t} 
 {D{^0}(t)\choose \overline D{^0}(t)} = 
\left({\hbox{\bf M}} - \frac{i}{2} {{\bf \Gamma}}\right) 
 {D{^0}(t)\choose \overline D{^0}(t)} \, ,
\label{eqn:schro}
\end{equation}
where the {\bf M} and ${\bf \Gamma}$ matrices are Hermitian, 
and $CPT$ invariance requires 
$M_{11}=M_{22}\equiv M$ and $\Gamma_{11}=\Gamma_{22}\equiv \Gamma$.
The off-diagonal elements of these matrices
describe the dispersive or long-distance and the absorptive
or short-distance contributions to $D^0$-$\DZB$ mixing.
We define the neutral $D$ meson mass eigenstates to be
\begin{eqnarray}
|D_1\rangle &=& p|D^0\rangle + q|\DZB\rangle \\
|D_2\rangle &=& p|D^0\rangle - q|\DZB\rangle,
\end{eqnarray}
where $|p|^2+|q|^2=1$, and, following Ref.~\cite{Gronau:2001nr},
$D_1$ is the $CP$-odd state,
and $D_2$ is the $CP$-even state, so that $CP|D^0\rangle = -|\DZB\rangle$.
The corresponding eigenvalues of the Hamiltonian are
\begin{equation}
\lambda_{1,2}\!\equiv\!M_{1,2}\!-\!\frac{i}{2} \Gamma_{1,2}\!=\!\left(M\!-\!\frac{i}{2} \Gamma \right)
\!\pm\!\frac{q}{p} \left(M_{12}\!-\!\frac{i}{2} \Gamma_{12} \right),
\label{eqn:eigenvalues}
\end{equation}
where $M_{1,2}$ and $\Gamma_{1,2}$ are the masses and decay widths,
respectively, and
\begin{equation}
\frac{q}{p} = 
\sqrt{\frac{M^*_{12}-\frac{i}{2}\Gamma^*_{12}}{M_{12}-\frac{i}{2}
\Gamma_{12}}} 
\,.
\label{alpha}
\end{equation}
A $\DZ$ can evolve into a $\DZB$ through on-shell intermediate
states, such as $K^+K^-$ with mass $M_{K^+K^-}\!=\!M_{\DZ}$, or through
off-shell intermediate states, such as those that might be present
due to new physics. 
This evolution through the former (latter) states is parametrized by
the dimensionless variables $-iy$ $(x)$.
We adopt the conventional definitions of these mixing parameters:
\begin{eqnarray}
x &\equiv& \frac{M_2-M_1}{\Gamma} \\
y &\equiv& \frac{\Gamma_2-\Gamma_1}{2\Gamma}.
\label{eqn:xandy}
\end{eqnarray}
The mixing probability, $R_M$, is approximately $(x^2+y^2)/2$~\cite{bib:asner}.
For hadronic flavored final states, the above time evolution is also governed
by the relative magnitudes and phases between Cabibbo-favored (CF) and
doubly-Cabibbo-suppressed (DCS) amplitudes, generically denoted by $r$ and
$\delta$, respectively.

Standard-Model-based predictions for $x$ and $y$, as well as a variety of
non-Standard-Model expectations, span several orders of
magnitude~\cite{hnncomp}.
Several non-Standard Models predict $|x| > 0.01$.  Contributions
to $x$ at this level could result from
the presence of new particles with masses as high as
100--1000~TeV~\cite{lns,ark}.
The Standard Model short-distance contribution to $x$ is determined by
the box diagram in which two virtual quarks and two virtual $W$ bosons
are exchanged. Next-to-leading order calculations show that the
short distance contributions to $x$ and $y$ are expected to be
comparable~\cite{Golowich:2005pt}. Long-distance effects are expected
to be larger but are difficult to estimate. It is likely that $x$ and $y$
contribute similarly to mixing in the Standard Model.

Measurement of the phase $\gamma/\phi_3$ of the Cabibbo-Kobayashi-Maskawa (CKM)
quark mixing matrix~\cite{ckm} is challenging and may
eventually be limited by experimental constraints on charm mixing~\cite{Silva:1999bd,Grossman:2005rp}. Several
methods have been proposed using $B^\mp \to D K^\mp$ decays: the 
Gronau-London-Wyler (GLW)~\cite{glw} method, where the $D$ decays to $CP$
eigenstates;
the Atwood-Dunietz-Soni (ADS)~\cite{ads} method, where the $D$ decays to flavor
eigenstates;
and the Dalitz-plot method~\cite{Giri:2003ty}, where the $D$ decays to a
three-body final state.
Uncertainties due to $D$ decays contribute to each of these methods.
The CLEO-c physics program~\cite{bib:cleoc} includes a variety of measurements
that will improve the determination of $\gamma/\phi_3$ from the $B$-factory
experiments, BABAR and Belle~\cite{Soffer:1998un,Bondar:2005ki}.
The pertinent components of this program are: improved constraints on charm 
mixing amplitudes (important for GLW), first measurement of the relative
strong phase $\delta_{K\pi}$ between $D^0$ and $\DZB$ decay to $K^+\pi^-$
(important for ADS), and studies of charm Dalitz plots tagged by hadronic
flavor or $CP$ eigenstates.
The total number of charm mesons accumulated at CLEO-c will be much smaller
than the samples already accumulated by the $B$-factories. However, 
the quantum correlation of the $D^0\DZB$ system near threshold provides
a unique laboratory in which to study charm.

The parameters $x$ and $y$ can be measured in a variety of ways.
The most precise constraints are obtained by exploiting the time-dependence of 
$D$ decays~\cite{bib:asner}. 
Previous attempts to measure $x$ and $y$ include: measurements of the
wrong-sign semileptonic branching ratio
$D^0\to K^+\ell^-\bar\nu_\ell$~\cite{e791sl,cleosl,babarsl,bellesl},
which is sensitive to $R_M$; decay rates to $CP$ eigenstates
$D^0\to K^+K^-$ and
$\pi^+\pi^-$~\cite{CLEOy,Belley,BABARy,FOCUSy,E791y,newBelley},
which are sensitive to $y$; the wrong-sign $D^0\to K^+\pi^-$ hadronic
branching ratio~\cite{cleokpi,BABARkpi,bellekpi,focuskpi}, which is sensitive
to $x^{\prime 2}\equiv(y\sin\delta_{K\pi}\!+\!
x\cos\delta_{K\pi})^2$ and $y^\prime\equiv y\cos\delta_{K\pi} -
x\sin\delta_{K\pi}$; 
wrong-sign $D^0\to K^+\pi^-\pi^0$~\cite{babarKpipi0} and
$D^0\to K^+\pi^-\pi^+\pi^-$~\cite{babarK3pi} decays;
and the decay rate of $\DZ \to K^0_S\pi^+\pi^-$\cite{cleokspipi},
which determines $\delta_{K^0_S\pi^+\pi^-}$ from a Dalitz-plot analysis and
measures $x$ and $y$.

Time-dependent analyses are not feasible at CLEO-c; however, previous
authors have found that the quantum-coherent $\DZ\DZB$
state~\cite{Kingsley:1975fe,Okun:1975di,Kingsley:1976cd} provides
time-integrated sensitivity, through the interference between amplitudes for
indistinguishable final states, to $y$ at ${\cal O}(1\%)$ and to
$\cos\delta_{K\pi}$ at ${\cal O}(0.1)$ in 1~fb$^{-1}$ of data at the
$\psi(3770)$~\cite{Gronau:2001nr,bib:cleoc,Atwood:2002ak}.  
In this paper, we extend the work of
Refs.~\cite{Gronau:2001nr,Atwood:2002ak,bigi,xing}
and develop a method for simultaneously measuring $x$, $y$, $r$, and $\delta$.
Unlike the proposed measurements of Ref.~\cite{Gronau:2001nr}, we do not rely
on external estimates of the relevant $D^0$ branching fractions.
Our method is a modified version of 
the double tagging technique originally developed to measure $D$ branching
fractions at the $\psi(3770)$~\cite{markiii-1,markiii-2,cleo-c}.  We make
use of rates for exclusive $D^0\DZB$ combinations, where both $D$ final states
are specified (known as double tags or DT), as well as inclusive rates,
where either the $D^0$ or $\DZB$ is identified and the other neutral $D$
decays generically (known as single tags or ST).
Although we estimate that CLEO-c will not have sufficient sensitivity to
observe Standard Model charm mixing (see Section~\ref{sec:experiment}), it
should be able to achieve a precision comparable to current experimental
results.
The analysis presented in this paper can also be applied to coherent
$K^0\KZB$, $B^0\BZB$, and $B_s^0\BZBS$ systems, although with some
additional complications.

\section{Formalism}\label{sec:formalism}

As in Refs.~\cite{Gronau:2001nr,Atwood:2002ak,bigi,xing}, we consider
the following categories of $D^0$ and $\DZB$ final states:
\begin{itemize}
\item $f$ or $\bar f$: hadronic states that can be reached from either $D^0$ or
$\DZB$ decay but that are not $CP$ eigenstates.  An example is $K^-\pi^+$,
which is produced via CF $D^0$ transitions or DCS $\DZB$ transitions.  We
include in this category Cabibbo-suppressed (CS) transitions as well as
self-conjugate final states of mixed $CP$, such as non-resonant
$K^0_S\pi^+\pi^-$.
\item $\ell^+$ or $\ell^-$: semileptonic or purely leptonic final states,
which, in the absence of mixing,
tag unambiguously the flavor of the parent $D$.
\item $S_+$ or $S_-$: $CP$-even and $CP$-odd eigenstates, respectively.
\end{itemize}

All $D^0$ decay modes can be treated uniformly if we enumerate charge-conjugate
final states separately, indexed by $j$ and
$\bar\jmath$.  For instance, $K^-\pi^+$ and $K^-\ell^+\nu_\ell$ are labeled
by $j$ and $K^+\pi^-$ and $K^+\ell^-\bar\nu_\ell$ by $\bar\jmath$.  $CP$
eigenstates appear in both lists because $\bar\jmath=j$.
We define the mode-dependent amplitude ratio 
$\langle j|\DZB\rangle/\langle j|D^0\rangle\equiv r_j e^{-i\delta_j}$, with
\begin{eqnarray}
\label{eq:rj} r_j &\equiv&
	\left|\frac{\langle j|\DZB\rangle}{\langle j|D^0\rangle}\right|\\
\label{eq:deltaj} -\delta_j &\equiv&
	\arg\left(\frac{\langle j|\DZB\rangle}
		{\langle j|D^0\rangle}\right)
	= \delta_{\rm strong} + \delta_{\rm weak} + \pi,
\end{eqnarray}
where the phase of $\pi$ results from our $CP$ convention. Since
$\delta_{\rm weak}$ in the charm sector is trivial (0 or $\pi$), $\delta_j$
corresponds to either $-\delta_{\rm strong}$ (if $j$ is CF) or
$\pi-\delta_{\rm strong}$ (if $j$ is CS).  Furthermore, if $CP$ is conserved,
then
$\langle j|\DZB\rangle/\langle j|D^0\rangle=\langle\bar\jmath| D^0\rangle/\langle\bar\jmath|\DZB\rangle$.
To resolve the ambiguity of whether to identify any given final state as
$j$ or $\bar\jmath$, we choose $0\leq r_j < 1$.

Since $x$ and $y$ have both been constrained to be less than
${\cal O}(1\%)$~\cite{pdg}, we generally keep terms only to leading order in
$x$ and $y$.
We denote decay amplitudes by $A_j \equiv \langle j|D^0\rangle$ and
$A_{\bar\jmath} \equiv \langle \bar\jmath|D^0\rangle$.
In our phase convention, $CP$ conjugate amplitudes are given by
\begin{eqnarray}
A_f &\equiv& \langle f|D^0\rangle = -\langle\bar f|\DZB\rangle \\
A_\ell &\equiv& \langle \ell^+|D^0\rangle = -\langle \ell^-|\DZB\rangle \\
A_{S_\pm} &\equiv& \langle S_\pm|D^0\rangle = \mp\langle S_\pm|\DZB\rangle.
\end{eqnarray}
We use a normalization in which $A_j^2$ is the $D^0\to j$ branching fraction
in the absence of mixing.  If mixing is present, then the branching fractions
for an isolated neutral $D$ meson produced in a $D^0$ or $\DZB$ flavor
eigenstate become~\cite{xing}
\begin{eqnarray}
{\cal B}_j &\equiv& {\cal B}(D^0\to j)\approx A_j^2(1+ r_j \tilde y_j) \\
{\cal B}_{\bar\jmath} &\equiv& {\cal B}(D^0\to\bar\jmath)\approx
	A_j^2(r_j^2 + r_j y'_j ) = {\cal B}_j R_j,
\end{eqnarray}
where $\tilde y_j \equiv y\cos\delta_j + x\sin\delta_j$,
$y'_j \equiv y\cos\delta_j - x\sin\delta_j$, and
$R_j\equiv \Gamma(\DZB\to j)/\Gamma(D^0\to j)\approx r_j^2 + r_j y'_j$.
The total $D^0$ decay rate is unaffected by mixing, so
\begin{equation}
\label{eq:totalRate}
	\sum_j \left(A_j^2 + A_{\bar\jmath}^2\right)
	= \sum_j A_j^2 \left( 1+ r_j^2\right)
	= \sum_j {\cal B}_j \left( 1+R_j \right)
	= 1.
\end{equation}
If the deviation of $q/p$ from unity is parametrized by two small
$CP$-violating parameters (magnitude and phase), then these parameters only
appear in products with $x$ and $y$; they can only modulate the strength of
the mixing signal.
Therefore, below, we assume $q/p=1$ and also that $CP$ is conserved
in the decay amplitudes ({\it i.e.},
$|\langle j|D^0\rangle|=|\langle\bar\jmath|\DZB\rangle|$), which allows
$y$ to be expressed in terms of $A_j^2$:
\begin{equation}
y = -\sum_j 2 A_j^2 r_j \cos\delta_f = \sum_{S_+} A_{S_+}^2 -
	\sum_{S_-} A_{S_-}^2 - \sum_f 2 A_f^2 r_f \cos\delta_f,
\end{equation}
where we have accounted for the fact that $S_\pm$ modes are simultaneously
labeled by $j$ and $\bar\jmath$.
Table~\ref{tab:rdelta} lists the values of $r_j$ and $\delta_j$ for each
final state category.  

\begin{table}[tb]
\caption{Values of the amplitude ratio magnitudes $r_j$ and phases $\delta_j$,
as well as uncorrelated branching fractions, for each final state category, to
first order in $x$ and $y$.}
\label{tab:rdelta}
\begin{center}
\begin{tabular}{c|ccc}
\hline\hline
~~~$j$~~~ & ~~~$r_j$~~~ & ~~~$\delta_j$~~~ & ~~~${\cal B}(D^0\to j)$~~~ \\
\hline
$f$ & $r_f$ & $\delta_f$ & $A_f^2(1+ r_f \tilde y_f)$ \\
$\bar f$ & $r_f$ & $\delta_f$ & $A_f^2(r_f^2+ r_f y'_f )$ \\
$\ell^+$ & 0 & --- & $A_\ell^2$ \\
$S_+$ & 1 & $\pi$ & $A_{S_+}^2 (1-y)$ \\
$S_-$ & 1 & 0 & $A_{S_-}^2 (1+y)$ \\
\hline\hline
\end{tabular}
\end{center}
\end{table}

As shown in Ref.~\cite{Goldhaber:1976fp}, a $D^0\DZB$ pair produced through
a virtual photon in the reaction $e^+e^-\to D^0\DZB+m\gamma+n\pi^0$ is in
a $C=(-1)^{m+1}$ state.  Thus, at the $\psi(3770)$, where no additional
fragmentation particles are produced, there is only $C$-odd,
while at higher energies above $D^*D$ threshold, we can access both $C$
eigenstates.  The DT rates for final states $j$ and $k$ are given
by~\cite{Gronau:2001nr,Atwood:2002ak,bigi,xing}
\begin{eqnarray}
\Gamma^{C-}(j, k) &=&
	Q_M\left|A^{(-)}(j,k)\right|^2 + R_M\left|B^{(-)}(j,k)\right|^2 \\
\nonumber
\Gamma^{C+}(j, k) &=&
	Q_M'\left|A^{(+)}(j,k)\right|^2 + R_M'\left|B^{(+)}(j,k)\right|^2 +
	C^{(+)}(j,k),
\end{eqnarray}
where
\begin{eqnarray}
\label{eq:apm}
A^{(\pm)}(j,k) &\equiv& \langle j|D^0\rangle\langle k|\DZB\rangle \pm
	\langle j|\DZB\rangle\langle k|D^0\rangle \\
\label{eq:bpm}
B^{(\pm)}(j,k) &\equiv& \frac{p}{q}\langle j|D^0\rangle\langle k|D^0\rangle \pm
	\frac{q}{p}\langle j|\DZB\rangle\langle k|\DZB\rangle \\
\label{eq:cpm}
C^{(+)}(j,k) &\equiv& 2\Re \left\{ A^{(+)*}(j,k)B^{(+)}(j,k)\left[
	\frac{y}{(1-y^2)^2} + \frac{ix}{(1+x^2)^2}\right]\right\}\\
Q_M &\equiv& \frac{1}{2}\left[\frac{1}{1-y^2}+\frac{1}{1+x^2}\right]
	\approx 1 - \frac{x^2-y^2}{2} \\
R_M &\equiv& \frac{1}{2}\left[\frac{1}{1-y^2}-\frac{1}{1+x^2}\right]
	\approx \frac{x^2+y^2}{2} \\
Q'_M &\equiv&
	\frac{1}{2}\left[\frac{1+y^2}{(1-y^2)^2}+\frac{1-x^2}{(1+x^2)^2}\right]
	\approx Q_M - x^2 + y^2 \\
R'_M &\equiv&
	\frac{1}{2}\left[\frac{1+y^2}{(1-y^2)^2}-\frac{1-x^2}{(1+x^2)^2}\right]
	\approx 3 R_M .
\end{eqnarray}
Using Equations~\ref{eq:rj} and~\ref{eq:deltaj}, we find
\begin{eqnarray}
|A^{(\pm)}(j, \bar k)|^2 \approx |B^{(\pm)}(j, k)|^2 &\approx&
	A_j^2 A_k^2 \left[ 1 + r_j^2 r_k^2 \pm r_j r_k v^-_{jk} \right] \\
|A^{(\pm)}(j, k)|^2 \approx |B^{(\pm)}(j, \bar k)|^2 &\approx&
	A_j^2 A_k^2 \left[ r_j^2 + r_k^2 \pm r_j r_k v^+_{jk} \right] \\
C^{(+)}(j, \bar k) &\approx& A_j^2 A_k^2 c^+_{jk} \\
C^{(+)}(j, k) &\approx& A_j^2 A_k^2 c^-_{jk},
\end{eqnarray}
where $z_j\equiv 2\cos\delta_j$, $w_j\equiv 2\sin\delta_j$,
$v^\pm_{jk} \equiv (z_j z_k \pm w_j w_k)/2$, and
\begin{equation}
c_{jk}^\pm \equiv
\frac{y}{(1-y^2)^2}\left[ (1+r_j^2) r_k z_k + (1+r_k^2) r_j z_j\right] \pm
\frac{x}{(1+x^2)^2}\left[ (1-r_j^2) r_k w_k + (1-r_k^2) r_j w_j\right].
\end{equation}

\begin{table}[tb]
\caption{$D^0\DZB$ DT branching fractions for modes containing $f$ or $\bar f$,
to leading order in $x$ and $y$.
}
\label{tab:DTRateFactors1}
\begin{center}
\begin{tabular}{c|cc}
\hline\hline
 &       $f$ & $\bar f$ \\
\hline
& \multicolumn{2}{c}{\boldmath $C=-1$} \\
$f$      &
	$A^4_f R_M \left[1+r_f^2(2-z_f^2)+r_f^4\right]$ \\
$\bar f$ &
	$A^4_f \left[1+r_f^2(2-z_f^2)+r_f^4\right]$ &
	$A^4_f R_M \left[1+r_f^2(2-z_f^2)+r_f^4\right]$ \\
$f'$ &
	$A^2_f A^2_{f'} \left(r_f^2 + r_{f'}^2 - r_f r_{f'} v^+_{ff'}\right)$ &
	$A^2_f A^2_{f'} \left(1 + r_f^2 r_{f'}^2-r_f r_{f'} v^-_{ff'}\right)$ \\
$\bar f'$ &
	$A^2_f A^2_{f'} \left(1 + r_f^2 r_{f'}^2-r_f r_{f'} v^-_{ff'}\right)$ &
	$A^2_f A^2_{f'} \left(r_f^2 + r_{f'}^2-r_f r_{f'} v^+_{ff'}\right)$ \\
$\ell^+$ &
	$A^2_f A^2_\ell r_f^2$ &
	$A^2_f A^2_\ell$ \\
$\ell^-$ &
	$A^2_f A^2_\ell$ &
	$A^2_f A^2_\ell r_f^2$ \\
$S_+$ &
	$A^2_f A^2_{S_+} \left[1 + r_f(r_f + z_f)\right]$ &
	$A^2_f A^2_{S_+} \left[1 + r_f(r_f + z_f)\right]$ \\
$S_-$ &
	$A^2_f A^2_{S_-} \left[1 + r_f(r_f - z_f)\right]$ &
	$A^2_f A^2_{S_-} \left[1 + r_f(r_f - z_f)\right]$ \\
\hline
& \multicolumn{2}{c}{\boldmath $C=+1$} \\
$f$      &
	$2 A^4_f r_f\left(r_f + y_f' + r_f^2\tilde y_f\right)$ \\
$\bar f$ &
	$A^4_f\left[ 1-r_f^2(2-z_f^2)+r_f^4+4r_f(\tilde y_f + r_f^2 y_f')\right]$ &
	$2 A^4_f r_f\left(r_f + y_f' + r_f^2\tilde y_f\right)$ \\
$f'$ &
	$A^2_f A^2_{f'} \left(r_f^2 + r_{f'}^2 + r_f r_{f'} v^+_{ff'} + 2c^-_{ff'}\right)$ &
	$A^2_f A^2_{f'} \left(1+r_f^2 r_{f'}^2 + r_f r_{f'} v^-_{ff'} + 2c^+_{ff'}\right)$ \\
$\bar f'$ &
	$A^2_f A^2_{f'} \left(1+r_f^2 r_{f'}^2 + r_f r_{f'} v^-_{ff'} + 2c^+_{ff'}\right)$ &
	$A^2_f A^2_{f'} \left(r_f^2 + r_{f'}^2 + r_f r_{f'} v^+_{ff'} + 2c^-_{ff'}\right)$ \\
$\ell^+$ &
	$A^2_f A^2_\ell \left(r_f^2 + 2 r_f y_f'\right)$ &
	$A^2_f A^2_\ell \left(1+2r_f\tilde y_f\right)$ \\
$\ell^-$ &
	$A^2_f A^2_\ell \left(1+2r_f\tilde y_f\right)$ &
	$A^2_f A^2_\ell \left(r_f^2 + 2 r_f y_f'\right)$ \\
$S_+$ &
	$A^2_f A^2_{S_+} \left[1 + r_f(r_f - z_f)\right](1-2y)$ &
	$A^2_f A^2_{S_+} \left[1 + r_f(r_f - z_f)\right](1-2y)$ \\
$S_-$ &
	$A^2_f A^2_{S_-} \left[1 + r_f(r_f + z_f)\right](1+2y)$ &
	$A^2_f A^2_{S_-} \left[1 + r_f(r_f + z_f)\right](1+2y)$ \\
\hline\hline
\end{tabular}
\end{center}
\end{table}

\begin{table}[tb]
\caption{$D^0\DZB$ DT branching fractions for semileptonic modes and $CP$
eigenstates, to leading order in $x$ and $y$.
}
\label{tab:DTRateFactors2}
\begin{center}
\begin{tabular}{c|cccc}
\hline\hline
&~~~~~~~~~~$\ell^+$~~~~~~~~~~ & ~~~~~~~~~~$\ell^-$~~~~~~~~~~ &
~~~~~~~~~~$S_+$~~~~~~~~~~ & ~~~~~~~~~~$S_-$~~~~~~~~~~ \\
\hline
& \multicolumn{4}{c}{\boldmath $C=-1$} \\
$\ell^+$ &
	$A^4_\ell R_M$ &
	&
	\\
$\ell^-$ &
	$A^4_\ell$ &
	$A^4_\ell R_M$ &
	\\
$S_+$ &
	$A^2_\ell A^2_{S_+}$ &
	$A^2_\ell A^2_{S_+}$ &
	0 \\
$S_-$ &
	$A^2_\ell A^2_{S_-}$ &
	$A^2_\ell A^2_{S_-}$ &
	$4 A^2_{S_+}A^2_{S_-}$ &
	0 \\
\hline
& \multicolumn{4}{c}{\boldmath $C=+1$} \\
$\ell^+$ &
	$3 A^4_\ell R_M$ &
	&
	\\
$\ell^-$ &
	$A^4_\ell$ &
	$3 A^4_\ell R_M$ &
	\\
$S_+$ &
	$A^2_\ell A^2_{S_+}(1 - 2y)$ &
	$A^2_\ell A^2_{S_+}(1 - 2y)$ &
	$2A^4_{S_+}(1-2y)$ \\
$S_-$ &
	$A^2_\ell A^2_{S_-}(1 + 2y)$ &
	$A^2_\ell A^2_{S_-}(1 + 2y)$ &
	0 &
	$2A^4_{S_-}(1+2y)$ \\
$S'_+$ &
	$A^2_\ell A^2_{S'_+}(1 - 2y)$ &
	$A^2_\ell A^2_{S'_+}(1 - 2y)$ &
	$4A^2_{S_+}A^2_{S'_+}(1-2y)$ & 0 \\
$S'_-$ &
	$A^2_\ell A^2_{S'_-}(1 + 2y)$ &
	$A^2_\ell A^2_{S'_-}(1 + 2y)$ &
	0 &
	$4A^2_{S_-}A^2_{S'_-}(1+2y)$ \\
\hline\hline
\end{tabular}
\end{center}
\end{table}

In Tables~\ref{tab:DTRateFactors1} and~\ref{tab:DTRateFactors2}, we give
the $D^0\DZB$ branching fractions to DT final states for 
$C$-odd and $C$-even initial states,
evaluated using the above formulae.
If both $D^0$ and $\DZB$ decay to the same final state, we divide the
$|A^{(\pm)}|^2$ and $C^{(+)}$ terms by 2.
In Table~\ref{tab:DTRateFactors2}, the entries with vanishing rate would be
non-zero only the presence of {\it both} mixing
and $CP$ violation~\cite{bigi,xing}.
If the $D^0\DZB$ decay were uncorrelated, these DT branching fractions would be
${\cal B}(j,\bar k)={\cal B}(\bar\jmath, k) = {\cal B}_j {\cal B}_k(1+R_j R_k)$
and
${\cal B}(j,k) = {\cal B}(\bar\jmath,\bar k) = {\cal B}_j{\cal B}_k(R_j+R_k)$,
with a factor of 1/2 if $j=k$ in the latter expression.

The $D^0\DZB$ inclusive branching fraction to the ST final state $j$ is
obtained by summing all DT branching fractions containing
$j$ and is found to be the same for $C$-odd and $C$-even (to first order in
$x$ and $y$) and simply related to the isolated $D^0$ branching fractions:
\begin{equation}
\label{eq:STSum}
{\cal B}(j, X) = \sum_k \left[ {\cal B}(j,k) + {\cal B}(j,\bar k) \right]
\approx A_j^2 \left[1 + r_j^2 + r_j z_j y \right] =
{\cal B}_j + {\cal B}_{\bar\jmath}.
\end{equation}
Table~\ref{tab:STRateFactors} shows these ST branching fractions evaluated for
the three categories of final states.

\begin{table}[tb]
\caption{$D^0\DZB$ inclusive ST branching fractions, to leading order in $x$
and $y$.}
\label{tab:STRateFactors}
\begin{center}
\begin{tabular}{c|c}
\hline\hline
$j$ & ~~~~~~~~~~$C=+1$ and $C=-1$~~~~~~~~~~ \\
\hline
$f$     & $A^2_f \left[1 + r_f^2 + r_f z_f y\right]$ \\
$\ell$  & $A^2_\ell$ \\
$S_\pm$ & $A^2_{S_\pm} (1\mp y)$ \\
\hline\hline
\end{tabular}
\end{center}
\end{table}

The total $D^0\DZB$ rate is obtained by summing either DT or ST rates:
\begin{equation}
\Gamma_{D^0\DZB} =
	\sum_{j, k\geq j}\left[\Gamma(j,k)+\Gamma(\bar\jmath,\bar k)\right]
	+ \sum_{j,k} \Gamma(j,\bar k) =
\frac{1}{2}\sum_j \left[ \Gamma(j,X)+\Gamma(\bar\jmath,X)\right].
\end{equation}
Like the total rate for an isolated $D^0$, the total $D^0\DZB$ rate is
unaffected by mixing and quantum correlations.

For the $C$-odd configuration, with only one mode of type $f$, the ST and DT
rates depend on only four independent parameters: $r_f$, $z_f$, $y$, and $R_M$.
For the $C$-even configuration, there is one additional parameter, $w_f x$,
which appears via $y'_f$ and $\tilde y_f$.
So, there is, in principle, sensitivity to $x$ from the $C$-even configuration,
although no information can be gained if $\delta_f$ is 0 or $\pi$.
In addition, estimates of $R_M$ and $y$ can be combined to obtain $x^2$.
However, in all of these cases, it is impossible, without knowing the
sign of $\delta_f$, to determine the sign of $x$.
The mixing and amplitude ratio parameters can be isolated by forming ratios
of DT rates and double ratios of ST rates to DT rates.  
Table~\ref{tab:doubleRatios} lists a selection of these ratios and functions
thereof, evaluated to leading order in $r_j^2$, $x$, and $y$.

\begin{table}[tb]
\caption{Selected ratios of DT rates and double ratios of ST rates to DT rates,
evaluated to leading order in $r_f^2$, $x$, and $y$.  Rates are represented by
the notation $\Gamma_{jk}\equiv \Gamma (j,k)$ and $\Gamma_j\equiv \Gamma(j,X)$.
}
\label{tab:doubleRatios}
\begin{center}
\begin{tabular}{c|cc}
\hline\hline
& ~~~~~~$C$-odd~~~~~~ & $C$-even\\
\hline
$(1/4)\cdot (\Gamma_{\ell S_+}\Gamma_{S_-}/\Gamma_{\ell S_-}\Gamma_{S_+} - \Gamma_{\ell S_-}\Gamma_{S_+}/\Gamma_{\ell S_+}\Gamma_{S_-})$ &
$y$ & $-y$ \\

$(\Gamma_{f\ell^-}/4 \Gamma_{f})\cdot (\Gamma_{S_-}/\Gamma_{\ell S_-} - \Gamma_{S_+}/\Gamma_{\ell S_+})$ &
$y$ & $-y$ \\

$(\Gamma_{f\bar f}/4\Gamma_{f})\cdot (\Gamma_{S_-}/\Gamma_{\bar f S_-} - \Gamma_{S_+}/\Gamma_{\bar f S_+})$ &
$y + r_f z_f$ & $-(y + r_f z_f)$ \\

$(\Gamma_{f}\Gamma_{S_+ S_-}/4)\cdot(1/\Gamma_{fS_-}\Gamma_{S_+}-1/\Gamma_{fS_+}\Gamma_{S_-})$ &
$y + r_f z_f$ & $0$ \\

$(\Gamma_{\bar f}/2)\cdot (\Gamma_{S_+ S_+}/\Gamma_{\bar f S_+}\Gamma_{S_+} - \Gamma_{S_- S_-}/\Gamma_{\bar f S_-}\Gamma_{S_-})$ &
$0$ & $y + r_f z_f$ \\

$\Gamma_{ff}/\Gamma_{f\bar f}$ &
$R_M$ &
$2 r_f^2 + r_f ( z_f y - w_f x )$ \\

$\Gamma_{f\ell^+}/\Gamma_{f\ell^-}$ &
$r_f^2$ &
$r_f^2 + r_f ( z_f y - w_f x )$ \\

$\Gamma_{\ell^\pm\ell^\pm}/\Gamma_{\ell^+\ell^-}$ &
$R_M$ &
$3 R_M$ \\
\hline\hline
\end{tabular}
\end{center}
\end{table}

\section{Effect on Branching Fraction Measurements}

If quantum correlations are ignored when using coherent $D^0\DZB$ pairs to
measure $D^0$ branching fractions, then biases may result.  For instance, if a
measured branching fraction, denoted by $\widetilde{\cal B}$, is obtained by
dividing reconstructed ST yields by the total number of $D^0\DZB$ pairs
(${\cal N}$), then
$\widetilde{\cal B}$ differs from the desired branching fraction, ${\cal B}$,
by the factors given in Table~\ref{tab:STRateFactors}.

Using a double tag technique pioneered by MARK III~\cite{markiii-1,markiii-2},
CLEO-c has recently measured
${\cal B}(D^0\to K^-\pi^+)$, ${\cal B}(D^0\to K^-\pi^+\pi^0)$, and
${\cal B}(D^0\to K^-\pi^+\pi^+\pi^-)$ in a self-normalizing way ({\it i.e.},
without knowledge of the luminosity or $D^0\DZB$ production cross
section)~\cite{cleo-c}, using $C$-odd $D^0\DZB$ pairs from the $\psi(3770)$.
Measured ST and DT yields and efficiencies are
combined in a least-squares fit~\cite{brfit} to extract the branching
fractions and ${\cal N}$.
Quantum correlations were not explicitly accounted for in this analysis, but
their effects were included in the systematic uncertainties.
Only flavored final states were considered, and DCS contributions to the ST
yields were removed, so the observed branching fractions are:
\begin{eqnarray}
\widetilde{\cal B}^{C-}_f \approx
	\frac{\Gamma^{C-}(f,\bar f')}{\Gamma^{C-}(\bar f', X)}
	&\approx& {\cal B}_f \left( 1 - r_f\tilde y_f - r_{f'}\tilde y_{f'} -
		r_f r_{f'} v_{ff'}^- \right) \\
&\approx& {\cal B}_f \left[ 1 - 2r_f\tilde y_f + r_f^2(2-z_f^2)\right]
\ {\rm for} \ f'=f.
\end{eqnarray}
Similarly, the relationship between the observed 
$\widetilde{\cal N}^{C-}$, which is used to obtain the $D^0\DZB$
cross section, and the desired ${\cal N}^{C-}$ is
\begin{eqnarray}
\widetilde{\cal N}^{C-} \approx
	\frac{\Gamma^{C-}(f,X)\Gamma^{C-}(\bar f',X)}
	{\Gamma^{C-}(f,\bar f')}
&\approx& {\cal N}^{C-} \left( 1 + r_f\tilde y_f + r_{f'}\tilde y_{f'} +
		r_f r_{f'} v_{ff'}^- \right)\\
&\approx& {\cal N}^{C-} \left[ 1 + 2r_f\tilde y_f - r_f^2(2-z_f^2)
	\right] \ {\rm for} \ f'=f.
\end{eqnarray}
The differences between the observed and the desired quantities are expected to
be ${\cal O}(1\%)$.

In principle, any $D^0$ branching fraction measured with DT yields in a
coherent $D^0\DZB$ system is subject to such considerations.
Analogous caveats pertain to $B^0$ and $B_s^0$ branching fractions measured
with coherent $B^0\BZB$ and $B_s^0\BZBS$ pairs.
However, for $K^0\KZB$ decays, such as those studied by KLOE, the situation
is generally simpler.  There, the desired branching
fractions~\cite{kloe1,kloe2} are for $K^0_S$ and $K^0_L$, rather than for $K^0$
and $\KZB$, so corrections for the lifetime asymmetry ($y\approx 0.997$) need
not be applied.

\section{Experimental Sensitivity}\label{sec:experiment}

The least-squares fit discussed in the previous section can be extended
to extract the parameters $y$, $x^2$, $r_f$, $z_f$, and $w_f x$
($C$-even only), in addition to ${\cal B}_j$ and ${\cal N}$.
Efficiency-corrected ST and DT yields are identified
with the functions given in Tables~\ref{tab:DTRateFactors1},
\ref{tab:DTRateFactors2}, and~\ref{tab:STRateFactors}.
We estimate uncertainties on the fit parameters based on approximately
$3\times 10^6$ $D^0\DZB$ pairs,
using efficiencies and background levels similar to those found at CLEO-c.
In the rate ratios in Table~\ref{tab:doubleRatios}, uncertainties that
are correlated by final state, such as tracking efficiency uncertainties,
cancel exactly.  Therefore, the uncertainties on the mixing and amplitude ratio
parameters stem primarily from statistics and from uncorrelated systematic
uncertainties.

The decay modes considered are listed in Table~\ref{tab:sensitivityModes}.
There exist, in principle, different $r_f$, $z_f$, and $w_f$ parameters for
each mode $f$ included in the fit.  Therefore, for simplicity, we
include only one such mode in the analysis: $D\to K^\pm\pi^\mp$.  In practice,
adding more hadronic flavored modes does not noticeably improve the precision
of $y$ because the limiting statistical uncertainty comes from DT yields
involving $S_\pm$.  The branching fraction determinations, however, would
benefit from having additional modes in the fit.

We omit ST yields for modes with a neutrino or a $K^0_L$,
which typically escapes detection, because they are difficult to measure.
In principle, one could reconstruct the remainder of the event inclusively
to infer the presence of the missing particle from energy and momentum
conservation~\cite{nurecon}.  The efficiency
of this method depends on the hermeticity of the detector.

For DT modes with one missing neutrino or $K^0_L$, this method is more
straightforward to implement because the $D^0$ and $\DZB$ are both
reconstucted exclusively.  Therefore, we do include these yields in the
fit.  In the case of a missing $K^0_L$, one must veto $K^0_S$ decays, which
would have the opposite $CP$ eigenvalue.

For DT modes with two undetected particles, one can constrain the event
kinematically, up to a twofold ambiguity~\cite{Brower:1997be}.  Background
events tend to fail these constraints, so the signal can be isolated
effectively.  We assume this method is used to measure $\ell^+\ell^-$,
$\ell^\pm\ell^\pm$, $K^0_L\ell^\pm$, and $K^0_LK^0_L$ DT yields.

\begin{table}[tb]
\caption{Final states included in the fits, along with assumed branching
fractions, signal efficiencies, and expected single tag yields for
${\cal N} = 3\times 10^6$.  ST yields in square brackets are not
included in any of the fits.}
\label{tab:sensitivityModes}
\begin{center}
\begin{tabular}{ccccc}
\hline\hline
Final State & ~~Type~~ & ~~${\cal B}$ (\%)~~ & ~~$\epsilon$ (\%)~~
	& ~~ST Yield ($10^3$) \\
\hline
$K^-\pi^+$ & $f$
	& 3.91
	& 66
	& 78 \\
$K^+\pi^-$ & $\bar f$
	& 3.91
	& 66
	& 78 \\
\hline
$K^- e^+ \nu_e$ & $\ell^+$
	& 3.5
	& 62
	& [65] \\
$K^+ e^- \bar\nu_e$ & $\ell^-$
	& 3.5
	& 62
	& [65] \\
\hline
$K^+ K^-$ & $S_+$
	& 0.389
	& 59
	& 14 \\
$\pi^+\pi^-$ & $S_+$
	& 0.138
	& 73
	& 6.0 \\
$K^0_S\pi^0\pi^0$ & $S_+$
	& 0.89
	& 15
	& 8.0 \\
$K^0_L\pi^0$ & $S_+$
	& 1.15
	& 62
	& [43] \\
$(K^0_S\pi^+\pi^-)_{CP+}$ & $S_+$
	& 1.0
	& 38
	& 27 \\
$(K^0_L\pi^+\pi^-)_{CP+}$ & $S_+$
	& 1.0
	& 76
	& [46] \\
\hline
$K^0_S\phi$ & $S_-$
	& 0.47
	& 7.7
	& 2.2 \\
$K^0_S\omega$ & $S_-$
	& 1.15
	& 14
	& 9.7 \\
$K^0_S\pi^0$ & $S_-$
	& 1.15
	& 31
	& 21 \\
$K^0_L\pi^0\pi^0$ & $S_-$
	& 0.89
	& 30
	& [16] \\
$(K^0_S\pi^+\pi^-)_{CP-}$ & $S_-$
	& 1.0
	& 38
	& 23 \\
$(K^0_L\pi^+\pi^-)_{CP-}$ & $S_-$
	& 1.0
	& 76
	& [46] \\
\hline\hline
\end{tabular}
\end{center}
\end{table}

The input ST yields are listed in Table~\ref{tab:sensitivityModes}, and
the input DT yields are derived from products of the ST branching fractions
and efficiencies.
In most cases, the background is negligible
($\sigma_N^{\rm stat}\approx \sqrt{N}$).  We also include a conservative 1\%
uncorrelated sytematic uncertainty on each yield measurement.
For modes that only have contributions from $R_M$, which we assume to be zero,
we use yield measurements of $0\pm 1\pm 1$.  Yields for forbidden modes
($S_\pm S_\pm$ for $C$-odd, $S_+ S_-$ for $C$-even) are not included.
The fit accounts for the statistical correlations among ST and DT yields.

We perform fits for both $C$ eigenvalues using equal numbers of $D^0\DZB$
pairs.  To improve the precision of these fits, we include 
external measurements of 
branching
fractions~\cite{pdg}.  
The first two columns of Table~\ref{tab:sensitivity}
show the expected uncertainties on the mixing and strong phase parameters
from these fits.  The dramatic difference between the $y$ uncertainties for the
two cases stems from the negative correlation between ${\cal B}_{S_+}$ and
${\cal B}_{S_-}$ introduced by the presence of $S_+S_-$ yields for $C$-odd;
these branching fractions are positively correlated for $C$-even.
From the second line of Table~\ref{tab:doubleRatios}, it can be seen that
a negative correlation increases the uncertainty on $y$.
The difference between the $x^2$
uncertainties for the two cases reflects the factor of 3 accompanying
$R_M$ in $\Gamma^{C+}(\ell^\pm,\ell^\pm)$.

In reality, $C$-even $D^0\DZB$ pairs [from $e^+e^-\to D^0\DZB\gamma(n\pi^0)$]
are produced above $D^*D$ threshold and
are more difficult to identify than $C$-odd $D^0\DZB$ pairs produced at
the $\psi(3770)$.  In particular, one must distinguish $D^0\DZB\gamma(n\pi^0)$,
which gives $C$-even, from $D^0\DZB(n\pi^0)$ and $D^0\DZB\gamma\gamma$, which
give $C$-odd.  While it is possible to do so for DT modes, where the entire
event is reconstructed, a large uncertainty is incurred for ST modes.
Therefore, we perform a third fit that combines yields from
both $C$ configurations, but with $C$-even ST yields omitted.  Because of the 
smaller cross section and
efficiencies for $D^0\DZB\gamma$ (only half the soft photons can be
identified), $C$-even DT yields above $D^*D$ threshold are an order of
magnitude smaller than $C$-odd yields from an equal luminosity at the
$\psi(3770)$.
Results for this fit are also shown in Table~\ref{tab:sensitivity}.

\begin{table}[tb]
\caption{Estimated uncertainties (statistical and systematic, respectively)
for different $C$ configurations, with 
branching fractions
constrained to the world averages.  We include $C$-even ST yields
in the second column, but not the third.}
\label{tab:sensitivity}
\begin{center}
\begin{tabular}{c|c|ccc}
\hline\hline
Parameter~ & ~Value~ & ~~~~~${\cal N}^{C-} = 3\times 10^6$~~ &
~~${\cal N}^{C+} = 3\times 10^6$~~~~~ &
${\cal N}^{C-} = 10\cdot{\cal N}^{C+} = 3\times 10^6$ \\
\hline
$y$
	& $0$
	& $\pm 0.015\pm 0.008$
	& $\pm 0.007\pm 0.003$
	& $\pm 0.012\pm 0.005$ \\
$x^2$ $(10^{-3})$
	& $0$
	& $\pm 0.6\pm 0.6$
	& $\pm 0.3\pm 0.3$
	& $\pm 0.6\pm 0.6$ \\
$\cos\delta_{K\pi}$
	& $1$
	& $\pm 0.21\pm 0.04$
	& $\pm 0.27\pm 0.05$
	& $\pm 0.20\pm 0.04$ \\
$x\sin\delta_{K\pi}$
	& $0$
	& ---
	& $\pm 0.022\pm 0.003$
	& $\pm 0.027\pm 0.005$ \\
$r^2$ $(10^{-3})$
	& $3.74$
	& $\pm 1.0\pm 0.0$
	& $\pm 1.7\pm 0.1$
	& $\pm 1.0\pm 0.0$ \\
\hline\hline
\end{tabular}
\end{center}
\end{table}

One important source of systematic uncertainty not included above is the
purity of the initial
$C=\pm 1$ state.  The sample composition can be determined from the ratios
\begin{equation}\label{eq:relativeCContributions}
\frac{\Gamma(S_+,S'_+)\Gamma(S_-,S'_-)}{\Gamma(S_+,S_-)\Gamma(S'_+,S'_-)} =
\frac{\Gamma(S_+,S'_+)\Gamma(S_-,S'_-)}{\Gamma(S_+,S'_-)\Gamma(S'_+,S_-)} =
\frac{4\Gamma(S_+,S_+)\Gamma(S_-,S_-)}{\Gamma^2(S_+,S_-)} =
\left(\frac{{\cal N}^{C+}}{{\cal N}^{C-}}\right)^2,
\end{equation}
assuming $CP$ is conserved.
Thus, if we include the forbidden $S_+S_+$, $S_-S_-$, and $S_+S_-$ DT yields
that were previously omitted, then we can construct every other ST
or DT yield as a sum of
$C$-odd and $C$-even contributions, with their relative sizes constrained by
Equation~\ref{eq:relativeCContributions}.  In this way, the systematic
uncertainty is absorbed into the statistical uncertainties, and the $C$
content of the sample is self-calibrating.  If the fit is
performed on pure samples, with either
$\Gamma(S_\pm,S_\pm)$ or $\Gamma(S_+,S_-)$ measured to be consistent with zero,
then the uncertainties given in Table~\ref{tab:sensitivity} suffer no
degradation.  On the other hand, if
${\cal N}^{C-} = {\cal N}^{C+}$, then there is an
unfortunate cancellation of terms containing $z_f$ or $y$,
and these parameters cannot be determined at all.

To demonstrate the importance of semileptonic modes in this analysis, we
consider a variation on the above fits.  If no semileptonic yields are
measured, then, for $C$-odd there is
only one independent combination of $r_f$, $z_f$, and $y$:
$[r_f z_f + y(1 + r_f^2)]/(1 + r_f^2 + r_f z_f y) \approx r_f z_f + y$.
On the other hand, if only the $\ell^\pm K^0_L$ DT modes are omitted, then $y$
and $z_f$ can be determined separately but with uncertainties
approximately 50\% larger than those in Table~\ref{tab:sensitivity}, where
the $\ell^\pm K^0_L$ DT modes are included.

The $\ell^\pm\ell^\pm$ and $\ell^+\ell^-$ DT modes improve the uncertainties
on $x^2$ and ${\cal B}_\ell$, but not on $y$ and $z_f$.  
However, if the self-calibrating fit described above were performed without
these modes, then $r_f$ would be strongly coupled to $x^2$ and
$z_f$ [through $\Gamma(f,f)$], resulting in inflated
uncertainties.  In order to stabilize the fit, it would be
necessary to fix the value of $x^2$. 

\begin{figure}[t]
\includegraphics*[width=.5\textwidth]{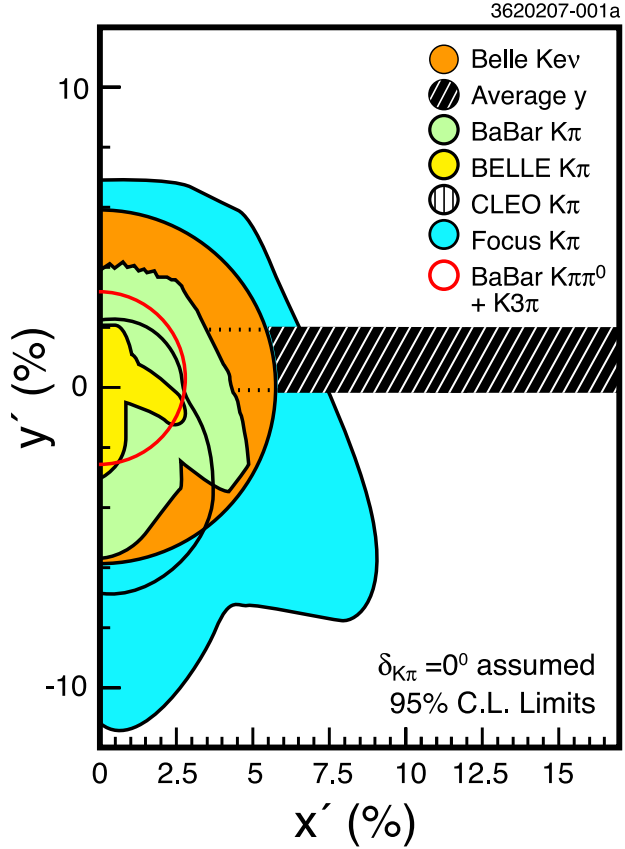}
\caption
{\label{fig:xpyp}
Allowed regions in the plane of $y_{K\pi}^\prime$ versus
$x_{K\pi}^\prime$~\cite{E791y, CLEOy, BABARy, newBelley, bellesl, cleokpi, BABARkpi, bellekpi, focuskpi, babarKpipi0, babarK3pi, cleokspipi}, assuming $\delta_{K\pi}=0$.
A non-zero value for $\delta_{K\pi}$ would rotate the
$\Delta\Gamma$
confidence region clockwise about the origin by an angle $\delta_{K\pi}$.
}
\end{figure}

\begin{figure}[t]
\includegraphics*[width=.5\textwidth]{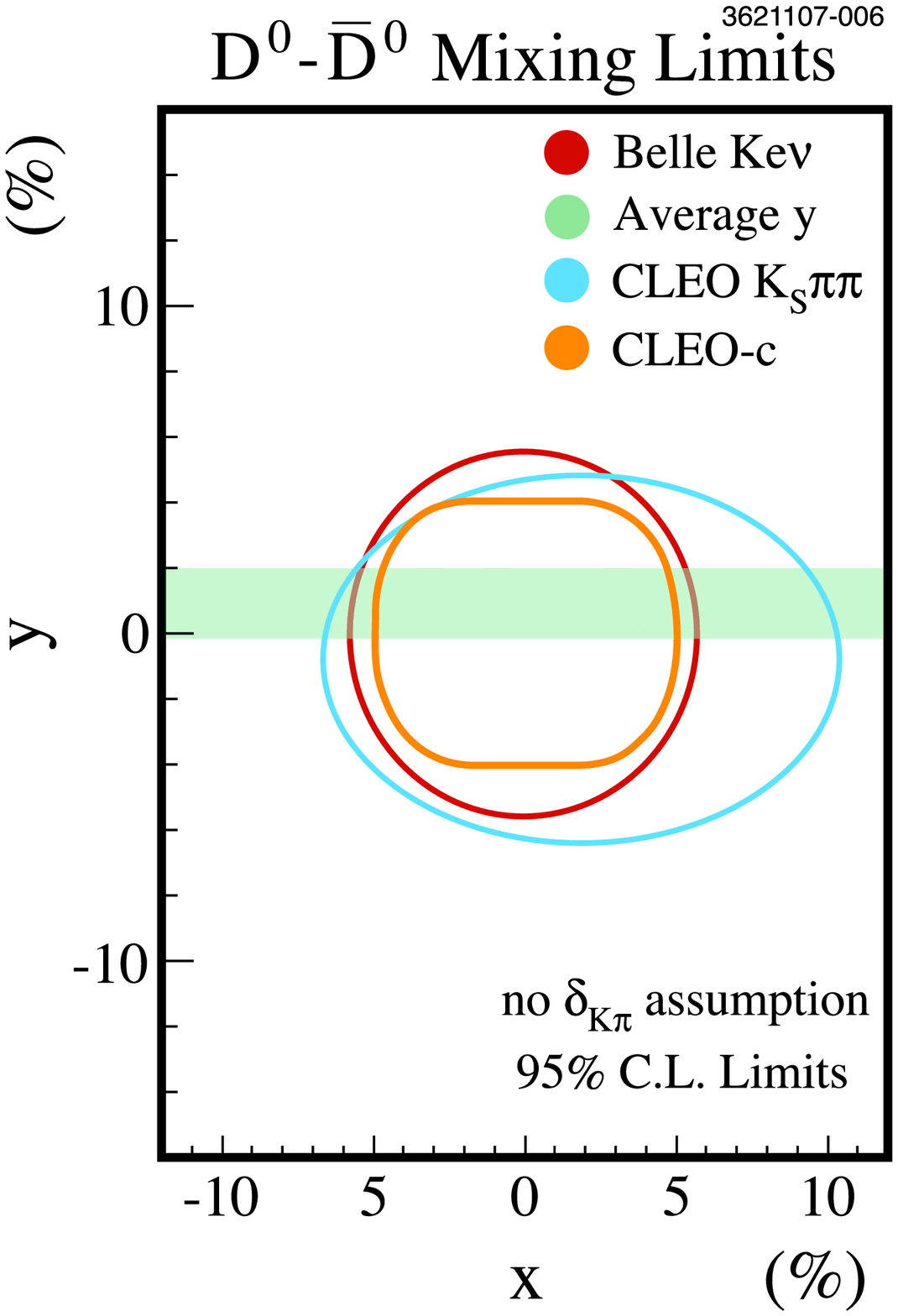}
\caption
{\label{fig:xy}
Allowed regions in the plane of $y$ versus $x$~\cite{E791y, CLEOy, BABARy, newBelley, bellesl, cleokspipi}, with our 95\% C.L. contour
for ${\cal N}^{C-} = 10\cdot{\cal N}^{C+} = 3\times 10^6$
superimposed.
}
\end{figure}

Experimental constraints on charm mixing are usually presented as a
two-dimensional region either in the plane of $x_{K\pi}^\prime$ versus
$y_{K\pi}^\prime$ or in the plane of
$x$ versus $y$. In Figures~\ref{fig:xpyp} and~\ref{fig:xy}, we compare current
constraints with those projected using the method described in this paper.
In Figure~\ref{fig:xpyp}, we show the results of
the time-dependent analyses of $\DZ \to K^+\pi^-$ from CLEO~\cite{cleokpi},
BABAR~\cite{BABARkpi}, 
Belle~\cite{bellekpi}, and FOCUS~\cite{focuskpi}. The regions for CLEO, BABAR,
and Belle allow for $CP$ violation in
the decay amplitude, in the mixing amplitude, and in the interference between
these two processes, while the FOCUS
result does not. Several experiments have also measured
$y$ directly by comparing the $D^0$ decay time for the
$K^-\pi^+$ final state to that for the $CP$ eigenstates $K^+K^-$ and
$\pi^+\pi^-$.
The allowed region for $y$ (labeled $\Delta\Gamma$)
shown in Figures~\ref{fig:xpyp} and~\ref{fig:xy} is the average of the
results from E791~\cite{E791y}, CLEO~\cite{CLEOy}, BABAR~\cite{BABARy}, and
Belle~\cite{newBelley}. 
In depicting the $y$ and Dalitz-plot results in
Figure~\ref{fig:xpyp}, we assume $\delta_{K\pi}=0$;
a non-zero value for $\delta_{K\pi}$ would rotate the
$\DZ \to K^-K^+/\pi^+\pi^-$ 
confidence region clockwise about the origin by an angle $\delta_{K\pi}$.
The best limit on $R_M$ from semileptonic searches for charm mixing
($\DZ \to \DZB \to K^+\ell^-\nu_\ell$), shown in 
Figures~\ref{fig:xpyp} and~\ref{fig:xy}, is from the Belle
experiment~\cite{bellesl}.
Figure~\ref{fig:xpyp} also displays the BABAR $R_M$ limits from
wrong-sign $D^0\to K^+\pi^-\pi^0$ and $D^0\to K^+\pi^-\pi^+\pi^-$.
Semileptonic results 
from E791~\cite{e791sl}, BABAR~\cite{babarsl}, and CLEO~\cite{cleosl} are not
shown. 

In Figure~\ref{fig:xy}, we plot the projected $95\%$ confidence
level (C.L.) contour for the fit in the third column of
Table~\ref{tab:sensitivity}, along with the results of a time-dependent
Dalitz-plot analysis of $\DZ \to K^0_S\pi^+\pi^-$ by CLEO~\cite{cleokspipi},
as well as the lifetime and semileptonic results discussed above.
We note that our sensitivity to $x$ depends strongly on the value of
$\delta_{K\pi}$, while our sensitivity to $y$ does not, and lowering the
value of $|\cos\delta_{K\pi}|$ would reduce the area of the contour.
In general, our expected upper limits compare favorably with the current best
limits on charm mixing.

\section{Summary}

We have derived ST and DT rate expressions for correlated $D^0\DZB$ pairs
in a definite $C$ eigenstate.  Interference between amplitudes for
indistinguishable final states enhances some $D^0$ decays and suppresses
others, depending on the $D^0$-$\DZB$ mixing parameters and on the magnitudes
and phases of various amplitude ratios.  By examining different types of final
states, which have different interference characteristics, we can extract
the mixing and amplitude ratio parameters, in addition to the branching
fractions.  In contrast to previous measurements of mixing
parameters~\cite{bib:asner}, our method is both time-independent and sensitive
to $x$ and $y$ at first order, so it is subject to different systematic
uncertainties.  Also, it is unique to threshold production where the $D^0\DZB$
initial state is known, unlike with $D^0$ mesons produced at fixed target
experiments or through $D^*$ decays, as at the $B$-factories and at LEP.

When performing this analysis for $K^0\KZB$, $B^0\BZB$, and $B_s^0\BZBS$
decays, one should incorporate $CP$ violation and non-trivial weak
phases.  In principle, given a source of coherent
$B_s^0\BZBS$ pairs, such as those produced through the $\Upsilon(5S)$,
$B_s^0$-$\BZBS$ mixing could be probed in a
fashion similar to that presented in this paper for $D^0$-$\DZB$ mixing.

\acknowledgments

We wish to thank David Cinabro, Qing He, Adam Lincoln, Peter Onyisi,
Ron Poling,
Alexander Scott, and Ed Thorndike for many helpful discussions
and for commenting on this manuscript.
This work was supported in part by the National Science Foundation under
Grant No. PHY-0202078.


\begin{thebibliography}{99}

\bibitem{gmp} M.~Gell-Mann and A.~Pais,
Phys. Rev. {\bf 97}, 1387 (1955).
\bibitem{goodetal}
R.H.~Good {\it et al.}, Phys.\ Rev.\ {\bf 124}, 1223 (1961).
\bibitem{glr} M.~K.~Gaillard, B.~W.~Lee, and J.~Rosner,
Rev. Mod. Phys. {\bf 47}, 277 (1975).
\bibitem{Albrecht}
H.~Albrecht {\it et al.},
Phys.\ Lett.\ B {\bf 192}, 245 (1987).
\bibitem{rosner87}
J.~L.~Rosner, in proceedings of
{\it Hadron 87 (2nd Int. Conf. on Hadron
Spectroscopy)} Tsukuba, Japan, April 16-18, 1987,
edited by Y.~Oyanagi, K.~Takamatsu, and T.~Tsuru, KEK, 1987, p. 395.

\bibitem{Gronau:2001nr}
  M.~Gronau, Y.~Grossman and J.~L.~Rosner,
  Phys.\ Lett.\ B {\bf 508}, 37 (2001).

\bibitem{bib:asner}
D.~Asner, \emph{$\DZ$-$\DZB$ Mixing} in Review of Particle Physics,
\emph{Phys. Lett.}, \textbf{B592}, 1 (2004).

\bibitem{hnncomp}

H.~N.~Nelson,
in {\it Proc. of the 19th Intl. Symp. on Photon and Lepton
  Interactions at High Energy LP99 } ed. J.A. Jaros and M.E. Peskin,
  SLAC (1999);
S.~Bianco, F.~L.~Fabbri, D.~Benson and I.~Bigi,
  Riv.\ Nuovo Cim.\  {\bf 26N7}, 1 (2003);
A.~A.~Petrov,
  eConf {\bf C030603}, MEC05 (2003);
I.~I.~Y.~Bigi and N.~G.~Uraltsev,
  Nucl.\ Phys.\ B {\bf 592}, 92 (2001);
Z.~Ligeti,
  AIP Conf.\ Proc.\  {\bf 618}, 298 (2002);
A.~F.~Falk, Y.~Grossman, Z.~Ligeti and A.~A.~Petrov,
  Phys.\ Rev.\ D {\bf 65}, 054034 (2002);
C.~K.~Chua and W.~S.~Hou,
  arXiv:hep-ph/0110106.

\bibitem{lns}
M.~Leurer, Y.~Nir and N.~Seiberg,
Nucl.\ Phys.\ B {\bf 420}, 468 (1994).

\bibitem{ark} 
N.~Arkani-Hamed, L.~Hall, D.~Smith and N.~Weiner,
Phys.\ Rev.\ D {\bf 61}, 116003 (2000).

\bibitem{Golowich:2005pt}
  E.~Golowich and A.~A.~Petrov,
  Phys.\ Lett.\ B {\bf 625}, 53 (2005).

\bibitem{ckm} M.~Kobayashi and T.~Maskawa, Prog. Theor. Phys.
{\bf 49}, 652 (1973).

\bibitem{Silva:1999bd}
  J.~P.~Silva and A.~Soffer,
  Phys.\ Rev.\ D {\bf 61}, 112001 (2000).

\bibitem{Grossman:2005rp}
  Y.~Grossman, A.~Soffer and J.~Zupan,
  Phys.\ Rev.\ D {\bf 72}, 031501 (2005).


\bibitem{glw}
M.~Gronau and D.~Wyler,
  Phys.\ Lett.\ B {\bf 265}, 172 (1991);
M.~Gronau and D.~London.,
  Phys.\ Lett.\ B {\bf 253}, 483 (1991).

\bibitem{ads}
D.~Atwood, I.~Dunietz and A.~Soni,
  Phys.\ Rev.\ Lett.\  {\bf 78}, 3257 (1997);
D.~Atwood, I.~Dunietz and A.~Soni,
  Phys.\ Rev.\ D {\bf 63}, 036005 (2001).

\bibitem{Giri:2003ty}
A.~Giri, Y.~Grossman, A.~Soffer and J.~Zupan,
Phys.\ Rev.\ D {\bf 68}, 054018 (2003).

\bibitem{bib:cleoc}
R.~A.~Briere {\it et~al.}, \emph{{\rm CLNS-01-1742}} (2001).

\bibitem{Soffer:1998un}
  A.~Soffer, \emph{{\rm CLNS-97-1533}} (1997),
  arXiv:hep-ex/9801018.

\bibitem{Bondar:2005ki}
   A.~Bondar and A.~Poluektov,
   Eur.\ Phys.\ J.\  C {\bf 47}, 347 (2006).


\bibitem{e791sl}
E.~M.~Aitala {\it et al.}  [E791 Collaboration],
  Phys.\ Rev.\ Lett.\  {\bf 77}, 2384 (1996).

\bibitem{babarsl}
B.~Aubert {\it et al.}  [BABAR Collaboration],
  Phys.\ Rev.\ D {\bf 70}, 091102 (2004).

\bibitem{cleosl}
C.~Cawlfield {\it et al.}  [CLEO Collaboration],
  Phys.\ Rev.\ D {\bf 71}, 077101 (2005).

\bibitem{bellesl}
   U.~Bitenc {\it et al.}  [Belle Collaboration],
   Phys.\ Rev.\  D {\bf 72}, 071101 (2005).

\bibitem{E791y}
E.~M.~Aitala {\it et al.}  [E791 Collaboration],
Phys.\ Rev.\ Lett.\  {\bf 83}, 32 (1999).

\bibitem{FOCUSy}
J.~M.~Link {\it et al.}  [FOCUS Collaboration],
Phys.\ Lett.\ B {\bf 485}, 62 (2000).

\bibitem{CLEOy}
S.~E.~Csorna {\it et al.}  [CLEO Collaboration],
Phys.\ Rev.\ D {\bf 65}, 092001 (2002).

\bibitem{Belley}
K.~Abe {\it et al.}  [Belle Collaboration],
Phys.\ Rev.\ Lett.\  {\bf 88}, 162001 (2002).

\bibitem{BABARy}
B.~Aubert {\it et al.}  [BABAR Collaboration],
Phys.\ Rev.\ Lett.\  {\bf 91}, 121801 (2003).

\bibitem{newBelley}
K.~Abe {\it et al.}  [Belle Collaboration],
  arXiv:hep-ex/0308034.

\bibitem{cleokpi} 
R.~Godang {\it et al.}  [CLEO Collaboration],
Phys.\ Rev.\ Lett.\  {\bf 84}, 5038 (2000).

\bibitem{BABARkpi}
B.~Aubert {\it et al.}  [BABAR Collaboration],
Phys.\ Rev.\ Lett.\  {\bf 91}, 171801 (2003).

\bibitem{bellekpi}
K.~Abe {\it et al.}  [Belle Collaboration],
Phys.\ Rev.\ Lett.\  {\bf 94}, 071801 (2005).

\bibitem{focuskpi}
J.~M.~Link {\it et al.} [FOCUS Collaboration], 
Phys.\ Lett.\ B {\bf 618}, 23 (2005).

\bibitem{babarKpipi0}
  B.~Aubert {\it et al.}  [BABAR Collaboration],
  Phys.\ Rev.\ Lett.\  {\bf 97}, 221803 (2006).

\bibitem{babarK3pi}
  B.~Aubert {\it et al.}  [BABAR Collaboration],
  arXiv:hep-ex/0607090.

\bibitem{cleokspipi}
D.~M.~Asner {\it et al.}  [CLEO Collaboration],
Phys.\ Rev.\ D {\bf 72}, 012001 (2005).

\bibitem{Kingsley:1975fe}
  R.~L.~Kingsley, S.~B.~Treiman, F.~Wilczek and A.~Zee,
  Phys.\ Rev.\  D {\bf 11}, 1919 (1975).

\bibitem{Okun:1975di}
  L.~B.~Okun, B.~M.~Pontecorvo and V.~I.~Zakharov,
  Lett.\ Nuovo Cim.\  {\bf 13}, 218 (1975).

\bibitem{Kingsley:1976cd}
  R.~L.~Kingsley,
  Phys.\ Lett.\  B {\bf 63}, 329 (1976).

\bibitem{Atwood:2002ak}
  D.~Atwood and A.~A.~Petrov,
  Phys.\ Rev.\ D {\bf 71}, 054032 (2005).

\bibitem{bigi}
  I.~I.~Y.~Bigi and A.~I.~Sanda,
  Phys.\ Lett.\  B {\bf 171}, 320 (1986);
 I.I. Bigi, in: {\em Proceed. of the Tau-Charm Workshop}, L.V. Beers (ed.),
SLAC-Report-343, 1989, p. 169.

\bibitem{xing}
  Z.~Z.~Xing,
  Phys.\ Rev.\ D {\bf 55}, 196 (1997).

\bibitem{markiii-1} R.~M.~Baltrusaitis {\it et al.} [MARK III Collaboration],
Phys.~Rev.~Lett.~{\bf 56}, 2140 (1986).

\bibitem{markiii-2}  
J.~Adler {\it et al.} [MARK III Collaboration], Phys.~Rev.~Lett.~{\bf 60}, 89
(1988).

\bibitem{cleo-c}
  Q.~He {\it et al.}  [CLEO Collaboration],
  Phys.\ Rev.\ Lett.\  {\bf 95}, 121801 (2005).

\bibitem{Goldhaber:1976fp}
  M.~Goldhaber and J.~L.~Rosner,
  Phys.\ Rev.\ D {\bf 15}, 1254 (1977).

\bibitem{pdg} Particle Data Group, S.~Eidelman {\it et al.},
Phys.~Lett.~B {\bf 592}, 1 (2004).

\bibitem{brfit} W.~M.~Sun, Nucl.\ Instrum.\ Meth.\ A {\bf 556}, 325 (2006).

\bibitem{kloe1}
  A.~Aloisio {\it et al.}  [KLOE Collaboration],
  Phys.\ Lett.\ B {\bf 535}, 37 (2002).

\bibitem{kloe2}
  A.~Aloisio {\it et al.}  [KLOE Collaboration],
  Phys.\ Lett.\ B {\bf 538}, 21 (2002).

\bibitem{nurecon}
  J.~P.~Alexander {\it et al.}  [CLEO Collaboration],
  Phys.\ Rev.\ Lett.\  {\bf 77}, 5000 (1996).

\bibitem{Brower:1997be}
  W.~S.~Brower and H.~P.~Paar,
  Nucl.\ Instrum.\ Meth.\ A {\bf 421}, 411 (1999).


\end{thebibliography}
\end{document}